\documentclass[preprint,showpacs,preprintnumbers,amsmath,amssymb]{revtex4}

\usepackage{graphicx}
\usepackage{dcolumn}
\usepackage{bm}

\begin{document}

\title{Casimir force for a scalar field in warped brane worlds}

\author{Rom\'an Linares$^1$}
\email{lirr@xanum.uam.mx}
\author{Hugo A. Morales-T\'ecotl$^1$}
\email{hugo@xanum.uam.mx}
\author{Omar Pedraza$^{1,2}$}
\email{opedrazao@ipn.mx}

\affiliation{$^1$Departamento de F\'{\i}sica, Universidad Aut\'onoma Metropolitana Iztapalapa,\\
San Rafael Atlixco 186, C.P. 09340, M\'exico D.F., M\'exico,}

\affiliation{ $^2$Centro de Investigació\'on en Ciencia Aplicada y Tecnolog\'{\i}a Avanzada,\\
\, Unidad Legaria, Instituto Polit\'ecnico Nacional, \, \\
Av. Legaria 694, C.P. 11500, M\'exico D.F., M\'exico.
}


\begin{abstract}
In looking for imprints of extra dimensions in brane world models
one usually builts these so that they are compatible with known low
energy physics and thus  focuses on high energy effects.
Nevertheless, just as submillimeter Newton's law tests probe the
mode structure of gravity other low energy tests might apply to
matter. As a model example, in this work we determine the 4D Casimir
force corresponding to a scalar field subject to Dirichlet boundary
conditions on two parallel planes lying within the single brane of a
Randall-Sundrum scenario extended by one compact extra dimension.
Using the Green's function method such a force picks the
contribution of each field mode as if it acted individually but with
a weight given by the square of the  mode wave functions on the
brane. In the low energy regime one regains the standard 4D Casimir
force that is associated to a zero mode in the massless case or to a
quasilocalized or resonant mode in the massive one whilst the effect
of the extra dimensions gets encoded as an additional term.

\end{abstract}

\maketitle

\section{\label{introduction}Introduction}

Considering extra spatial dimensions making our observable 4D
universe a subspace of a higher dimensional spacetime has a long
tradition that started with the works of G. N\"ordstrom, T. Kaluza
and O. Klein. (see e.g. \cite{Appelquist:1987nr} and references
therein). Amongst the main motivations for such approaches we find
attempts to unify fundamental interactions, in particular including
gravity through Kaluza-Klein theories \cite{Appelquist:1987nr},
supergravities and string/M-theory \cite{Polchinski:1998rq}. On the
other hand, it has also been proposed extra dimensions may help to
come to terms with the cosmological constant and the hierarchy
problems
\cite{Akama:1982jy,Rubakov:1983bb,Rubakov:1983bz,Visser:1985qm,Arkani-Hamed:1998rs,Antoniadis:1998ig,
Arkani-Hamed:1998nn,Randall:1999ee,Randall:1999vf}. The current
status of the brane world idea, as it became popular to call the
field, can be seen in some reviews
\cite{Rubakov:2001kp,Feruglio:2004zf,Maartens:2003tw,PerezLorenzana:2005iv}
providing also some remarkable phenomenological aspects going from
astrophysics to cosmology. Most of these assume known low energy
physics remains unaltered thus focusing in the high energy regime.
However, just as Newton's law tests at sub-millimeter scale have
allowed to probe brane world scenarios
\cite{Hoyle:2000cv,Kapner:2006si} it is of interest to consider
other precise low energy experiments. Since physics in presence of
extra dimensions is linked to the mode structure of matter or
gravity fields a natural candidate test to study is the Casimir
force (other possibilities are high precision atomic experiments
\cite{Bluhm:2000tv,Luo:2006ck,MoralesTecotl:2006eh}).

 In 1948 H. B. G. Casimir predicted that two uncharged perfectly conducting flat plates, placed in
vacuum and separated by a distance $l$, should attract each other
with a force per unit area $A$ given by
$F(l)/A=-\frac{\pi^2}{240}\frac{\hbar c}{l^4}=-1.3\times
10^{-27}\frac{\mathrm{Nm}{^2}}{l^4}$. This force is a purely quantum
effect caused by the alteration of the electromagnetic field modes
due to the plates and it is described by QED \cite{Casimir:1948dh}.
We will refer to it as the standard or 4D Casimir force since it is
obtained in Minkowski spacetime. Although the effect is very weak it
becomes measurable when $l\sim 1 \mu$m. Indeed it has been
convincingly demonstrated by many experiments over the years
\cite{Sparnaay:1958wg,Mohideen:1998iz,Bressi:2002fr,Decca:2005yk,Klimchitskaya:2005df}
and its measurements have reached very high precision (see
\cite{Lamoreaux:2005gf} for a review of the current experimental
situation); also the Casimir force is convenient to consider
experimentally in that it implies macroscopic bodies as opposed to
atomic size systems.

Over the years the Casimir effect has been extended to different
fields, geometries, materials and models (see \cite{Milton:2001yy}
and references therein for a review of the Casimir effect in
different contexts). In general, it may be defined as the stress on
the bounding surface when a quantum field in vacuum state is
confined to a finite volume of space. In any case, the boundaries
restrict the modes of the quantum field giving rise to a force which
can be either attractive or repulsive, depending on the field and
the type of boundaries.

In particular, the Casimir effect has received a great deal of
attention within spacetime models including extra spatial
dimensions. For example, it has been discussed in the context of
string theory
\cite{Fabinger:2000jd,Gies:2003cv,Brevik:2000fs,Hadasz:1999tr}. Also
in the Randall-Sundrum model, the Casimir effect has been considered
to stabilize the separation between branes (radion)
\cite{Elizalde:2002dd,Garriga:2002vf,Pujolas:2001um,Flachi:2001pq,Goldberger:2000dv}
as well as within inflationary brane world universe models
\cite{Nojiri:2000bz}.

It is noteworthy some of these models can also affect the standard
4D Casimir force. A simple way to see this is to consider a field
defined on a higher dimensional scenario and then extracting a 4D
effective dynamics for it. The Casimir force is computed in 4D with
dispersion relations modified by the presence of the extra
dimensions, say
\begin{equation}\label{DispertionRel}
\frac{\omega^2}{c^2}=k^2+ \Delta k^2_{\mathrm{extra}}\,,
\end{equation}
where $\omega$ is the frequency and $k:=|\vec k|$ is the magnitude
of the wave vector of the mode. Usually $\Delta
k^2_{\mathrm{extra}}$ includes parameters of the higher dimensional
model and by demanding agreement of the corresponding Casimir force
with experiments it is either possible to set bounds for the
parameters or limit the phenomenological viability of the model.
Within this approach, the 4D Casimir force between parallel plates
has been computed for a scalar field in the presence of one
compactified universal extra dimension
\cite{Poppenhaeger:2003es,Cheng:2006rs,Pascoal:2007uh}, for the
effective 4D QED \cite{Linares:2005cj} that comes from a
Nielsen-Olesen vortex solution of the abelian Higgs model with
fermions coupled to gravity in 6D \cite{Randjbar-Daemi:2003qd} and
for a massless scalar field in the Randall-Sundrum models
\cite{Frank:2007jb}. In all these models extra dimensions produce
different kind of contributions to the dispersion relations. In the
first case the scenario is 5D, has topology $M_4\times S^1$ and the
second term in Eq. (\ref{DispertionRel}) consists of tower of
Kaluza-Klein massive modes of the scalar field, $\Delta
k^2_{\mathrm{extra}}=n^2/R^2$, where $n \in Z$ and $R$ is the radius
of $S^1$ \cite{Poppenhaeger:2003es,Cheng:2006rs,Pascoal:2007uh}. In
the second case, there are two extra dimensions that contribute to
the dispersion relations of the electromagnetic modes near the core
of the vortex that represents our 4D world. A continuous one
associated with a radial extra dimension and a discrete one
corresponding to an angular coordinate labeled by a vortex number,
$n_v \in N$. Explicitly, $\Delta k^2_{\mathrm{extra}} \approx
k_r^2+1/n_v^2 \ell ^2$. Here $\ell$ is a length scale defined by the
ratio of the 6D Newton constant and the 6D gauge coupling
\cite{Randjbar-Daemi:2003qd,Linares:2005cj}. The resulting 4D
Casimir force is in conflict with experiments, thus reducing the
phenomenological viability of the model. For a massless scalar field
in the 5D RSI scenario which includes two 3-branes separated by a
compact dimension, the contribution to the dispersion relation is a
tower of Kaluza-Klein modes exponentially suppressed, $\Delta
k_{\mathrm{extra}}^2\approx \kappa^2(n+1/4)^2 \, e^{-2\pi\kappa r}$.
$\kappa$ is the brane tension and $r$ is the separation distance
between the branes. A massless scalar field in the RSII model
including a single brane, yields a tower of continuous Kaluza-Klein
massive scalar modes, $\Delta k_{\mathrm{extra}}^2=m^2$, $m\geq 0$.
Upon correcting by the polarization in higher dimensions to go from
a massless scalar field to an electromagnetic one,  the experiments
imply an upper limit to $\kappa r$ for RSI. In the RSII case the
effect seems to be too small to be probed by experiment
\cite{Frank:2007jb}.

Implicit in the previous analysis is the assumption that the
massless scalar field Casimir force can be translated into the
electromagnetic one which is the one that is actually tested
experimentally. For this to be the case both scalar and
electromagnetic fields should have zero modes localized to our
brane. This holds for some but not all of the above scenarios.
Moreover, whereas in 4D Minkowski spacetime different methods yield
the same Casimir force it is not obvious whether the same results
hold for the effective models of brane world scenarios considered so
far. As a first step in this direction in this paper we compute the
4D Casimir force for a scalar field coming from a 6D scenario
RSII-1, consisting of a single 3-brane and 1 additional compact
extra dimension using the Green's function method, as opposed to the
dimensional regularization of previous analysis. RSII-1 owns the
non-trivial property of localizing gauge fields
\cite{Dubovsky:2000am}.

A salient feature of the modes corresponding to the non compact
dimension is linked to whether the scalar field has a 6D mass or
not. The massive case does not contain a zero mode and there are not
true localized modes with $m\neq 0$ but one that is quasilocalized.
In contrast the massless case has a spectrum incorporating a zero
mode with a continuum of massive modes
\cite{Bajc:1999mh,Giddings:2000mu}. This specific relation between
the mass spectrum and the bulk mass of the field is a characteristic
intrinsic to a noncompact dimension not shared by models containing
compact extra dimensions only \cite{Dubovsky:2000am}.

The structure of the paper is as follows. Section \ref{EDmodels}
sketches some basic features of warped and Kaluza-Klein models.
Special attention is payed to RSII and RSII-1. In section
\ref{6Dmassive} we present our analysis of the Casimir force for the
massive scalar field whereas section  \ref{masslessstringlike}
contains the massless case. Finally we discuss our results in
section \ref{secdiscussion}. Unless otherwise stated we use units in
which $\hbar=1,c=1$.

\section{Extra spatial dimensions}\label{EDmodels}

An important issue in considering higher dimensional scenarios is
the mechanism by which extra dimensions are hidden, in such a way
that the space-time is effectively 4D. There are two different ways
to implement this idea depending on whether the extra dimensions are
either {\it compact} or {\it noncompact}. Both possibilities can be
accommodated by means of the following $(4+d)$ dimensional metric
which is consistent with Poincar\'e invariance in 4D
\begin{equation}\label{genmetric}
ds^2_{4+d}=\sigma (x^c)g_{\mu \nu}(x^\rho)dx^\mu
dx^\nu-\gamma_{ab}(x^c)dx^a dx^b.
\end{equation}
Here Greek indices denote the usual 4D coordinates whereas Latin
indices denote extra dimensions, therefore in (\ref{genmetric}),
$g_{\mu \nu}$ is the metric of our world while $\gamma_{ab}$ is the
metric associated with the $d$ extra dimensions.

The first possibility arises in Kaluza-Klein type theories
\cite{Appelquist:1987nr}. Within this approach the $(4+d)$
space-time manifold is assumed to be separable in the form
$M_{4+d}=M_4 \times M_d$, where $M_4$ is our 4D world and $M_d$ is
the manifold associated with the small extra dimensions which are
{\it compact} and essentially homogeneous. The metric
(\ref{genmetric}) describes this possibility by taking $\sigma
(x^a)=1$, implying that $M_4$ is described by a factorizable
geometry, independent of $x^a$. Compactness of extra dimensions
ensures that space-time is effectively 4D at distances exceeding the
compactification scale $R$. This conclusion arises because from the
4D point of view, every multi-dimensional field (matter, gravity and
gauge fields) corresponds to a Kaluza-Klein tower of particles with
increasing masses. At low energies $E<R^{-1}$, only massless
particles can be produced, whereas at $E \sim R^{-1}$, the tower of
massive states is manifest and extra dimensions show up. Since
experimentally the Kaluza-Klein massive states have not been
observed, the energy scale $R^{-1}$ must be at least in the TeV
range, so in the Kaluza-Klein models, the size of the extra
dimensions must be microscopic $R \leq 10^{-17}$ cm.
\cite{Arkani-Hamed:1998rs,Antoniadis:1998ig}. The 4D effective
Casimir effect coming from a 5D massless scalar field in this
geometry has been discussed in
\cite{Poppenhaeger:2003es,Cheng:2006rs,Pascoal:2007uh}.

The second possibility considers {\it noncompact} extra dimensions
but still unobservable at low energies. There exist basically two
ways to obtain noncompact extra dimensions. One way is to consider
the metric (\ref{genmetric}) where $\sigma(x^a)$ is a conformal
factor depending on the extra coordinates only, implying that the
metric is non-factorizable, i.e., it does not correspond to a
product of $M_4$ and a manifold of extra dimensions. It was proposed
for first time as a space-time Ansatz to solve Einstein equations
with a positive cosmological constant in 6D \cite{Rubakov:1983bz}.
The second way is to identify our 4D world with the internal space
of a topological defect residing in a higher dimensional space-time
\cite{Rubakov:1983bb,Akama:1982jy,Visser:1985qm}, for instance, a
domain-wall in 5D, a string in 6D, a monopole in 7D, etc.
Generically all these types of backgrounds admit localization of
both fermionic and scalar field massless zero modes which are
associated with the 4D particles that we observe. It has also been
established that gravity can be localized on several topological
defect backgrounds. For instance, it was realized in
\cite{Randall:1999vf} that gravity can be localized on a 3-brane
(domain-wall), with positive tension and located at $y=0$ and
embedded in a 5D space-time whose metric is given by  two patches of
the symmetric space $AdS_5$  of radius $\kappa^{-1}$ and has the
structure of equation (\ref{genmetric}), namely,
\begin{equation}\label{metricRSIIin}
ds^2_{4+1}=e^{-2\kappa|y|} n_{\mu \nu}dx^\mu dx^\nu-dy^2.
\end{equation}
Here the extra dimension $y$ is noncompact and the parameter
$\kappa$ is determined by the 5D Planck mass and bulk cosmological
constant. This metric obeys the full 5D Einstein equations with
negative cosmological constant and the model is known as
Randall-Sundrum II model (RSII). One important property of the
geometry (\ref{metricRSIIin}) is that every field in this background
can be decomposed in 4D plane waves, due to its 4D Poincar\'{e}
invariance
\begin{equation}
\phi \propto \exp(ip_\mu x^\mu)\phi_p(z),
\end{equation}
where the 4-momentum $p_\mu$ coincides with the physical momentum on
the brane ($p^2=m^2$). The key point for the localization of gravity
is that a normalizable graviton zero mode ($m^2=0$) residing on the
domain-wall reproduces 4D gravity, while the continuum massive
spectrum of 5D gravitons living on the bulk, gives only a small
correction to the Newton's law at large distances. It is worth to
mention that there are other models in the noncompact extra
dimensions approach that also localize gravity (see e.g.
\cite{Feruglio:2004zf,PerezLorenzana:2005iv} and references therein
for a review of them). If one or more extra dimensions are infinite,
one naturally expects that particles may eventually leave the brane
and escape into the extra dimensions. In the RSII model, this
process is possible for gravitons \cite{Rubakov:2001kp}. If other
fields have bulk modes, the corresponding particles may also leave
the brane. As an example, fermions bound to the brane, even in the
absence of gravity, are capable of leaving the brane provided they
are given enough energy. From the 4D point of view this process
would show up as a process in which the charge is not conserved, for
example $e^- \rightarrow$ nothing. It is remarkable that an AdS
metric allows for such a process  at low enough energies
\cite{Dubovsky:2000av}. On the other hand the effective 4D Casimir
effect produced by a 5D massless scalar field in the geometry
(\ref{metricRSIIin}) has been recently considered in
\cite{Frank:2007jb} using the dimensional regularization technique.

Now in order to make the whole construction realistic it is also
important to have localization of gauge fields. There are several
scenarios that achieve this goal \cite{Rubakov:2001kp}, and in this
paper we are interested in the one that describes a 6D geometry with
a compact warped additional dimension. The metric, that we refer to
as RSII-1, away from the brane is
\begin{equation}\label{ecu:model}
ds^2_{4+2}=e^{-2\kappa |y|}\left(\eta_{\mu \nu}
dx^{\mu}dx^{\nu}-R^2d\theta^2\right)-dy^2 ,
\end{equation}
where $\theta$ is a compact extra coordinate taking values in the
interval $[0, 2 \pi)$, and $y$ is the coordinate along a single
non-compact extra dimension. This metric can be obtained in two
different ways, either, as an asymptotic solution to the 6D Einstein
equations with negative bulk cosmological constant and a 3-brane
(local string defect) with an appropriately tuned energy-momentum
tensor \cite{Gherghetta:2000qi,Gherghetta:2000jf}, or as in the RSII
model \cite{Randall:1999vf}, considering a codimension one brane (a
4-brane) with both positive tension and one compact dimension,
embedded in a 6D space-time. In this case the metric
(\ref{ecu:model}) obeys the full Einstein equations with essentially
the same fine-tuning condition between the tension of the brane and
the negative bulk cosmological constant as in the RSII model.
Gravity is localized on the brane because there exists a graviton
zero mode which is independent of $\theta$ outside the brane and
decrease at large $y$ as $e^{-2\kappa|y|}$, in complete analogy with
the RSII model. As in the Kaluza-Klein picture, the compact
dimension $\theta$, is invisible at low energies $E<R^{-1}$.

In this geometry it is possibility to localize not only spin 0 and
spin 2 fields, but also spin 1 fields
\cite{Oda:2000zc,Dubovsky:2000av}. This result is in contrast with
the RSII model for which is not possible to localize gauge fields
\cite{Bajc:1999mh}. The key point to have the localization is that
at low energies, $E<<R^{-1}$, the relevant gauge field
configurations are independent of $\theta$ and there exists a zero
mode gauge field also independent of $y$, which corresponds to a
massless vector boson localized on the string-like defect.

Localization of the gauge field besides gravity is the
characteristic that makes attractive the geometry (\ref{ecu:model}).
To have a complete picture of the model, it should be pointed out
that if we want to localize also particles of spin 1/2 or 3/2, it is
necessary to introduce additional interactions. Generalizations of
the metric (\ref{ecu:model}) to metrics with more than two extra
dimensions that also localize gauge fields can be found in
\cite{Dubovsky:2000av}. In this case a $(3+n)$-brane with $n$
compact coordinates is embedded in a $(3+n+2)$-dimensional
space-time,
\begin{equation}\label{RSIIN}
ds_{5+n}^{\,
2}=e^{-2\kappa|y|}\bigg[\eta_{\mu\nu}dx^{\mu}dx^{\nu}-\sum_{i=1}^nR_i^{2}d\theta_i^{2}\bigg]
-dy^2 \,.
\end{equation}
We denote this metric as RSII-$p$. In this paper we shall restrict
ourselves to the study of RSII-1 and we will only comment in the
discussion section some aspects of RSII-$p$ relevant to the
effective 4D Casimir force.

\section{Massive scalar field}\label{6Dmassive}

In this section we obtain the Casimir force for a massive scalar
field in the background metric RSII-1, eq. (\ref{ecu:model}). We
start by computing both the eigenfunctions and eigenvalues of the
differential operators for each independent coordinate that is
associated to the scalar field equation. With them we go on then to
compute the corresponding Green's function which is used to
determine the effective Casimir force.

Let us consider a massive scalar field $\Phi$ described by the
action in 6D
\begin{equation}\label{action6D}
S=\int R d \theta \, dy\, d{}\,^4x \,  \sqrt{-g} \left(
\frac{1}{2}\, g^{MN}\partial_M \Phi \, \partial_N \Phi - \frac{1}{2}
\, m_6^2 \, \Phi^2 \right),
\end{equation}
where the metric $g_{MN}$ is given by (\ref{ecu:model}). Here $m_6$
is the 6D mass of $\Phi$.

The field equation for $\Phi$ in the background (\ref{ecu:model})
hence becomes
\begin{equation}\label{ecu:field}
e^{2\kappa|y|}\Box_4\Phi-\frac{e^{2\kappa|y|}}{R^2}\partial_{\theta}^2\Phi
-\frac{1}{\sqrt{-g}}\partial_{y}\left[\sqrt{-g}
\partial_{y}\Phi\right]+m_6^2\Phi=0.
\end{equation}
$\Box_4$ here stands for the flat 4D Dalambertian corresponding to
$\eta_{\mu\nu}$. Assuming solutions of the form
$\Phi(x,\theta,y)=\varphi(x)\Theta(\theta)\psi(y)$ and performing
separation of variables it is straightforward to obtain the three
differential equations
\begin{eqnarray}
\left(\partial_\theta^{\, 2}+ m_\theta^{\,2} R^{\,2}\right) \Theta (\theta)& = & 0,\label{Thetaeq} \\
\left(\partial_{y}^{2}-5\kappa\,sgn(y)\partial_{y}-m_6^2+m^2\,e^{2\kappa|y|}\right)\psi(y)
& = & 0, \label{Yeq}\\
\left(\Box_4 + m_\theta^{\,2}+m^2 \right)\varphi (x)& = &
0,\label{varphieq}
\end{eqnarray}
where $m_\theta$ and $m$ are separation constants with units of
mass. From the 4D point of view eq. (\ref{varphieq}) corresponds to
an effective massive scalar field $\varphi(x)$ whose mass,
$m_4^2:=m_\theta^2+m^2$, picks up two independent contributions
corresponding to the compact and non compact extra dimensions,
respectively. Our first task is to determine this mass spectrum and
the associated eigenfunctions.

\subsection{Mode decomposition}

In this subsection we focus on the $\theta,y$ dependence, eqs.
(\ref{Thetaeq},\ref{Yeq}), whereas next one contains the $x$
dependence, eq. (\ref{varphieq}), which accounts for the Dirichlet
boundary conditions on flat planes. Now, in eq. (\ref{Thetaeq}),
$\Theta$ is subject to periodic boundary conditions so we obtain the
well known eigenfunctions and eigenvalues for a particle in a circle
\begin{equation}\label{SolTheta}
\Theta_n=\frac{1}{\sqrt{2\pi R}}e^{in\theta} \hspace{0.5cm}
\mbox{where} \hspace{0.5cm} n=m_\theta R \in {\mathbb{Z}}.
\end{equation}
Therefore the contribution of the extra compact coordinate to $m_4$,
is given in terms of the discrete spectrum, $m_\theta^2(n)=n^2/R^2$.
Hence lower dimensional physics is associated to the massless mode
$n=0$.  For $n \neq 0$ we have an infinite tower of discrete massive
modes as manifestation of the extra warped compact dimension which,
however, are considered to be suppressed in the low energy regime as
long as $R^{-1}\ll \kappa$. It must be stressed that eq.
(\ref{Thetaeq}) does not depend of $m_6$ so its solutions will hold
in the massless scalar field case.

As for the $y$ dependence, eq. (\ref{Yeq}), the question arises of
whether there is a localized scalar field mode on the brane. As it
will be shown below the key observation to answer this question is
that for a massive $\Phi$ a zero (massless) mode is precluded by
RSII-1 and there are not true localized modes with $m\neq 0$ neither
\cite{Dubovsky:2000am}. The best we can hope then is a
quasilocalized mode, which happens to be actually the case as it is
argued below. In contrast, as it is shown in section
\ref{masslessstringlike}, a massless $\Phi$ has an spectrum
incorporating a zero mode and a Kaluza-Klein continuum
\cite{Bajc:1999mh,Giddings:2000mu}.


Lacking a zero mode the next best thing we can have is a
quasilocalized or metastable mode. The corresponding state is
associated with a complex eigenvalue. There are several ways to
prove the existence of a metastable state (see
\cite{Dubovsky:2000am} for a detailed discussion). One of them is to
solve the equation (\ref{Yeq}) imposing the radiation boundary
conditions at $y\rightarrow \pm \infty$. It turns out that the
solutions are linear combinations of the Hankel functions $H_\gamma
^{(1)}(me^{\pm \kappa y}/\kappa)$. Requiring continuity of these
solutions and its derivatives on the brane one arrives at an
eigenvalue equation for $m$. The Hankel functions can be expanded in
the regime $m \ll \kappa$ and assuming additionally that $m_6 \ll
\kappa $, it is possible to show that
\begin{equation}
m=m_q-i\Delta,
\end{equation}
is a solution to the eigenvalue equation with
\begin{equation}
m_q^2=\frac{3}{5}\, m_6^2, \hspace{1cm} \mbox{and}
\hspace{1cm}\frac{\Delta}{m_q}=\frac{1}{6}\left(
\frac{m_q}{\kappa}\right)^3.
\end{equation}
Such a state can decay into the continuum modes and the physical
interpretation from the point of view of a 4D observer is that the
state corresponds to massive particle propagating in 3 spatial
dimensions for some time, and then disappears into the $y$ direction
\cite{Dubovsky:2000av}. Notice that the width $\Delta$ is suppressed
with respect to the mass $m_q$ by a factor $(m_q/\kappa)^3$ at small
$m_q/\kappa$.

\begin{figure}
\includegraphics[height=6cm]{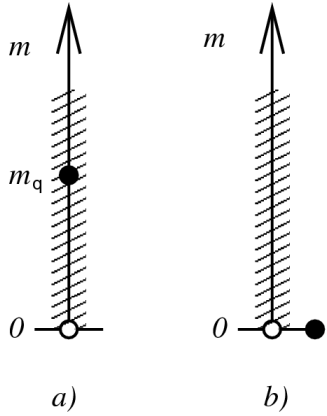}
\caption{\label{fig:modos} The figure shows the mass spectrum
corresponding to eq. (\ref{Yeq}). a) For $m_6\neq 0$ the spectrum is
continuous but it does not include the zero mode. $m_q$ represents
the mass of a quasilocalized mode. b) For $m_6=0$ the spectrum is
continuous and it does include the value $m=0$.}
\end{figure}

To proceed with our analysis it is convenient to have explicitly the
eigenfunctions for the modes. To do so let us observe eq.
(\ref{Yeq}) is invariant under reflection in $y$ and therefore it is
enough to solve the equation in the region $y>0$. Performing the
change of variable $\tilde y=e^{\kappa y}/\kappa$ and redefining the
function $\psi$ as $ \tilde \psi(\tilde y)=\tilde y^{5/2}\psi(\tilde
y)$, we obtain
\begin{equation}\label{ecforPsi}
\partial_{\tilde y}^2\tilde \psi+\frac{1}{\tilde y}\partial_{\tilde y}\tilde \psi+
\left(m^2-\frac{\gamma^2}{\tilde y^2}\right)\tilde \psi=0.
\end{equation}
Here the parameter $\gamma$ is given by
$\gamma^2=\left(\frac{5}{2}\right)^2+\left(\frac{m_6}{\kappa}\right)^2$.
When $m=0$, eq. (\ref{ecforPsi}) does not admit a solution
consistent with the corresponding boundary conditions. Thus there is
no zero mode. However, for $m>0$, the normalized modes are
\begin{equation}
\psi_m(y)=e^{\frac{5}{2}\kappa y}\,\sqrt{\frac{m}{2\kappa}}\left[
a_m J_{\gamma}\left(\frac{me^{\kappa y}}{\kappa}\right)+ b_m
N_{\gamma}\left(\frac{me^{\kappa y}}{\kappa}\right)  \right].
\end{equation}
The coefficients $a_m, b_m$ can be obtained from both the
normalization condition and the Neumann boundary condition (the
latter follwing from the reflection invariance of eq. (\ref{Yeq}))
\begin{eqnarray}
\int_{-\infty}^{\infty}dy\,e^{-3\kappa |y|}\,\psi_m
\psi_{m'}&=&\delta(m-m') \hspace{0.3cm} \nonumber\\
&\Rightarrow&  a_m^2+b_m^2=1, \label{normalizationmy}\\
\partial_y \psi_m(y)|_{y=0}&=&0 \label{Neumann},
\end{eqnarray}
where the weight factor in the measure (\ref{normalizationmy}) comes
from the Sturm-Liouville form of (\ref{Yeq}). The resulting
expressions for $a_m$ and $b_m$ are
\begin{equation}
a_m=-\frac{A_m}{\sqrt{1+A_m^2}},\quad b_m=\frac{1}{\sqrt{1+A_m^2}},
\end{equation}
where
\begin{equation}
A_m=\frac{N_{\gamma-1}\left(\frac{m}{\kappa}\right)-\left(\gamma-\frac{5}{2}\right)\frac{\kappa}{m}N_{\gamma}\left(\frac{m}{\kappa}\right)
}{
J_{\gamma-1}\left(\frac{m}{\kappa}\right)-\left(\gamma-\frac{5}{2}\right)\frac{\kappa}{m}J_{\gamma}\left(\frac{m}{\kappa}\right)
}.
\end{equation}
These states are not localized modes and therefore cannot represent
scalar particles in 4D. However we are interested in the low energy
regime and only modes with $m\ll \kappa$ are relevant, making it
possible to expand Bessel functions at small arguments. Assuming
additionally a light $\Phi$, namely $m_6\ll \kappa$, the
coefficients $A_m$ become
\begin{equation}
A_m\approx -\frac{2\Gamma(\gamma + 1)\Gamma(\gamma -1)}{\pi \left
(\gamma +
\frac{5}{2}\right)}\left(\frac{m}{2\kappa}\right)^{2-2\gamma}\left(1-2(\gamma
-1)\left(\gamma -
\frac{5}{2}\right)\left(\frac{\kappa}{m}\right)^2\right),
\end{equation}
and the squared wave functions of the modes at the brane behave like
\begin{equation}
\psi_m^2(y\rightarrow 0)\approx
\frac{9}{\pi}\left(\frac{m}{\kappa}\right)^{-4}\frac{1}{1+A_m^2}.
\end{equation}
There are two relevant mass regimes: one that turns out to
correspond to a quasilocalized mode, a regime where $\psi_m^2$ is
peaked, or equivalently where the $A_m$ are small,
\begin{equation}
A_m\approx \frac{1}{\Delta} (m_q-m).
\end{equation}
The other regime is defined by the light modes, $m \ll m_6$, for
which
\begin{equation}
1+A_m^2\approx \left( \frac{9}{5}\, \frac{\kappa^3
m_6^2}{m^5}\right)^2.
\end{equation}
Explicitly we have
\begin{equation}
\psi_m^2(0)\approx \left\{
\begin{array}{cc}\label{phiappox}
\frac{3}{2} \kappa   \, \delta(m_q-m) & \mbox{for}\, \, \, \, m \sim m_q ,  \\
\frac{m^6}{\pi \kappa^2 m_q^4} & \mbox{for}\, \, \, \, m\ll m_q .
\end{array} \right.
\end{equation}
An illustrative representation of this idea is presented in Fig.
\ref{fig:modos}.

Once we have obtained the mass spectrum of the scalar field modes by
looking at the $\theta, y$ dependence, our next task is to solve eq.
(\ref{varphieq}) for the $x$ dependence and incorporate them all to
get the Green's function.

\subsection{Green's function}
We want to apply Green's function formalism to compute the Casimir
force and hence, given the field equation for $\Phi$, eq.
(\ref{ecu:field}), the corresponding Green's function, $G_{6D}$ ,
should fulfill
\begin{equation}\label{green1}
 \left(e^{2\kappa|y|}\left[\Box_4-\frac{1}{R^2}
\partial_{\theta}^2\right]-\frac{1}{\sqrt{-g}}\partial_{y}\left[\sqrt{-g}\partial_{y}\,\,\right]+
m_6^2\right)G_{6D}=\frac{\delta({x}-{x}')\delta(R\theta-R\theta')\delta(y-y')}{\sqrt{-g}}.
\end{equation}
This can be expressed in terms of the eigenfunctions of the
differential operators for the different coordinates. In the
previous subsection we have presented the modes $\Theta_n, \psi_m$
accounting for the $\theta,y$ dependence, respectively, so we still
have to solve (\ref{varphieq}) depending on the 4D $x$ coordinates.
Now, since our aim is to compute the Casimir effect for the scalar
field in the standard setting of parallel plates, which in our case
lie along the brane, it is convenient to split up the 3D position
vector $\vec{x}$ in the way $\vec{x}=(\vec{x}_{\perp},z)$, where $z$
denotes the coordinate orthogonal to the plates and
$\vec{x}_{\perp}$ denotes a 2D vector orthogonal to the $z$
coordinate. Eq. (\ref{varphieq}) has to be solved in two different
regions, between planes ($0<z<l$) and to the right of the plane
$z=l$ (or equivalently to the left of the plane $z=0$) . Upon using
such a parametrization eq. (\ref{varphieq}) becomes
\begin{equation}\label{1DequationForz}
\left(-\partial_z^2 - \lambda^2  \right)\varphi (z) = 0,
\end{equation}
where the following definitions have been made
\begin{equation}\label{phitimesplanewave}
\varphi(x)=\varphi(t,\vec{x}_{\perp},z)=:\varphi(z)e^{-i\omega
t+i\vec{k}_{\perp}\cdot\vec{x}_{\perp}}\,,
\end{equation}
\begin{equation}\label{defLambda}
\lambda^2:= \omega^2-\vec{k}_{\perp}^2-\frac{n^2}{R^2}-m^2.
\end{equation}

The Green's function including only the information of the $z$
coordinate, say $G(z,z')$, is defined through eq.
(\ref{1DequationForz}), that is to say
\begin{equation}\label{GreenForz}
\left(-\partial_z^2 - \lambda^2  \right)G (z,z') = \delta(z-z').
\end{equation}
 $G(z,z')$ is usually termed reduced Green's function.

For the region  between plates ($0\leq z,z' \leq l$) eq.
(\ref{GreenForz}) is solved subject to the boundary conditions
\begin{equation}\label{boundaryGreen6D}
G(0,z')=G(l,z')=0,
\end{equation}
obtaining
\begin{equation}\label{ReducedGreenIn}
G_{\mathrm{in}}(z,z')=-\frac{1}{\lambda \sin \lambda \, l}\sin
\lambda \, z_< \sin \lambda(z_>-l),
\end{equation}
where $z_>(z_<)$ represents the greater (lesser) of $z$ and $z'$.

For the region to the right of the plates  $(l\leq z,z')$ the
solution is
\begin{equation}\label{ReducedGreenOut}
G_{\mathrm{out}}(z,z')=\frac{1}{\lambda}\left( \sin
{\lambda}(z_<-l)e^{i {\lambda}(z_>-l)} \right),
\end{equation}
which vanishes at $z=l$ and has outgoing boundary conditions as
$z\to\infty$,  $g(z,z')\sim e^{ik z}$. Notice that this solution is
different to the free Green's function, where plates are not
present.

Going back to the full Green's function and using the eigenfunction
expansion we have
\begin{equation}
G_{6D}=\sum_{n} \int \frac{dm}{\kappa} \int
\frac{d\omega}{2\pi}\frac{d^2k_{\perp}}{(2\pi)^2} \frac{1}{2\pi
R}e^{-i\omega
(t-t')+i\vec{k}_{\perp}\cdot(\vec{x}_{\perp}-\vec{x}_{\perp}')}e^{in(\theta-\theta
')}\psi_{ m}(y)\psi_{m}(y')G(z,z'), \label{GreenCommune}
\end{equation}
where $k_{\perp}$ is defined also transversal to the $z$ direction
and  $dm/\kappa$ is the proper measure on the continuum states
\cite{Randall:1999vf}. Also it is understood that $G(z,z')$ is
either $G_{\mathrm{in}}(z,z')$ or $G_{\mathrm{out}}(z,z')$ depending
on whether the region of interest is between or outside the plates.

Amusingly, eq. (\ref{GreenCommune}) can be rewritten as
\begin{equation}\label{6DComGreen}
G_{6D}=\sum_{n} \int \frac{dm}{\kappa}\, \Theta_n(\theta) \Theta_n
(\theta')\psi_{ m}(y)\psi_{m}(y') G_{4D}\left(x,x';m_4 \right),
\end{equation}
where $G_{4D}$ is the standard Green's function in 4D for a massive
scalar field of mass $m_4^2=\frac{n^2}{R^2}+m^2$ (see e.g.
\cite{Milton:2001yy}),
\begin{equation}
G_{4D}\left(x,x';m_4\right):=\int
\frac{d\omega}{2\pi}\frac{d^2k_{\perp}}{(2\pi)^2} e^{-i\omega
(t-t')+i\vec{k}_{\perp}\cdot(\vec{x}_{\perp}-\vec{x}_{\perp}')}G(z,z').
\end{equation}
The interpretation of the 6D Green's function (\ref{GreenCommune})
is a nice one: it is just a combination of 4D Green's functions
individual massive modes would produce irrespectively of whether
they are discrete or continuous weighted by the extra dimensional
wave functions' modes.

We can push the analytic calculation in the low energy regime by
considering light modes approximation $m\ll R^{-1}, \kappa$ and
therefore only the zero mode for the compact dimension, $n=0$, will
be considered. Also, since we shall calculate the Casimir force
between plates lying along the brane we restrict our analysis to
$y,\, y' \rightarrow 0$. The effective 4D Green's function in this
approximation splits into two pieces: one for quasilocalized, or
resonant, mode and the other for the light modes contribution
proper, as follows
\begin{equation}\label{Green'sFContr6DMassive}
G_{\mathrm{eff}}(x,x') = \frac{1}{2\pi R}\int_{m\, \sim \, m_q}
\frac{dm}{\kappa}\, \psi_{m_q}^2(0)\, G_{4D}(x,x';m) + \frac{1}{2\pi
R} \int_{m \, \ll \, m_6 \, \ll \kappa} \frac{dm}{\kappa} \,
\psi_m^2(0) \, G_{4D}(x,x';m)\,.
\end{equation}
Since the first term includes the factor $\psi_{m_q}^2(0)$ which is
a resonant state peaked at $m_q$, in light of (\ref{phiappox}), it
yields a contribution having the form of the standard 4D Green's
function of a massive scalar field whose mass is $m_q$. We will show
below it gives rise to the standard 4D Casimir force. The second
term is the contribution due to the rest of light massive modes.

\subsection{Casimir force}\label{seccasimir}

In order to calculate the Casimir force between plates from the
Green's function one uses the stress tensor \cite{Milton:2001yy}.
Let us focus on the plate located at $z=l$. This can be done
evaluating the discontinuity in the flux of the stress tensor across
the plate, i.e., we have to take into account the discontinuity of
the normal-normal component of the stress tensor
\begin{equation}\label{eqforce}
F=\int_0^A dx_{\perp}\int_{-\infty}^{\infty} dy \, e^{-3\kappa |y|}
\int_0^{2\pi}  R\, d\theta \left[ \langle
T_{zz}\rangle_{in}\bigg|_{z=l}\right. -\left. \langle
T_{zz}\rangle_{out} \bigg|_{z=l}\right].
\end{equation}
Here $A$ is the area of the plate.

For our massive scalar field the stress tensor is given by
\begin{equation}
T_{MN}=\partial_{M}\Phi\partial_{N}\Phi-\frac{1}{2}g_{MN}\partial^{P}\Phi\partial_{P}\Phi
+\frac{1}{2}m_6^2g_{MN}\Phi^2,
\end{equation}
and its vacuum expectation value may be obtained applying a
differential operator to the expression of the Green's function in
terms of the vacuum expectation value of the time ordering product
of two scalar fields
\begin{equation}
G(x,\theta,y;x',\theta',y')=i \left\langle T
[\Phi(x,\theta,y)\Phi(x',\theta',y')]\right\rangle .
\end{equation}
Computing the normal-normal component of the stress tensor to the
left of the plate, the obtained result in $z=l$ is
\begin{equation}\label{t1}
\langle T_{zz}\rangle_{\mathrm{in}}\bigg|_{z=l} \sim
\frac{1}{2i}\partial_{z}\partial_{z'}G_{in}(z,z')\bigg|_{z\rightarrow
z'= l} = \frac{i}{2} \lambda \cot \lambda \, l.
\end{equation}
To the right of the plate we obtain
\begin{equation}\label{t2}
\langle T_{zz}\rangle_{\mathrm{out}}\bigg|_{z=l}\sim
\frac{1}{2i}\partial_z\partial_{z'}G_{out}(z,z')\bigg|_{z\rightarrow
z'=l}=\frac{{\lambda}}{2}.
\end{equation}
Combining (\ref{6DComGreen}), (\ref{t1}) and (\ref{t2}) with
(\ref{eqforce}) gives the 4D force per unit area of the plates
(recall $y,y'\rightarrow 0$ on the brane)
\begin{equation}\label{fT}
f_{T}= \sum_{n}\int_0^{\infty}\frac{dm}{\kappa}\,\psi_m^2(0)
f_{4D}\left(m_4\right)\,,
\end{equation}
with $f_{4D}\left(m_4\right)$ is the standard 4D Casimir force for a
4D massive scalar field whose mass is $m_4^2=\frac{n^2}{R^2}+m^2$,
i.e. \cite{Milton:2001yy}
\begin{equation}\label{fCasimirStan}
f_{4D}(m_4)=\int\frac{d\omega
d^2k_{\perp}}{(2\pi)^3}\left(\frac{i}{2} \lambda \cot \lambda \, l
-\frac{{\lambda}}{2} \right)=-\frac{1}{32\,
\pi^{2}l^{4}}\int_{2lm_4}^{\infty} dx \,
\frac{x^2\sqrt{x^2-4l^2m^2_4}}{e^x-1}\, .
\end{equation}

\begin{figure}
\includegraphics[height=12cm,angle=-90]{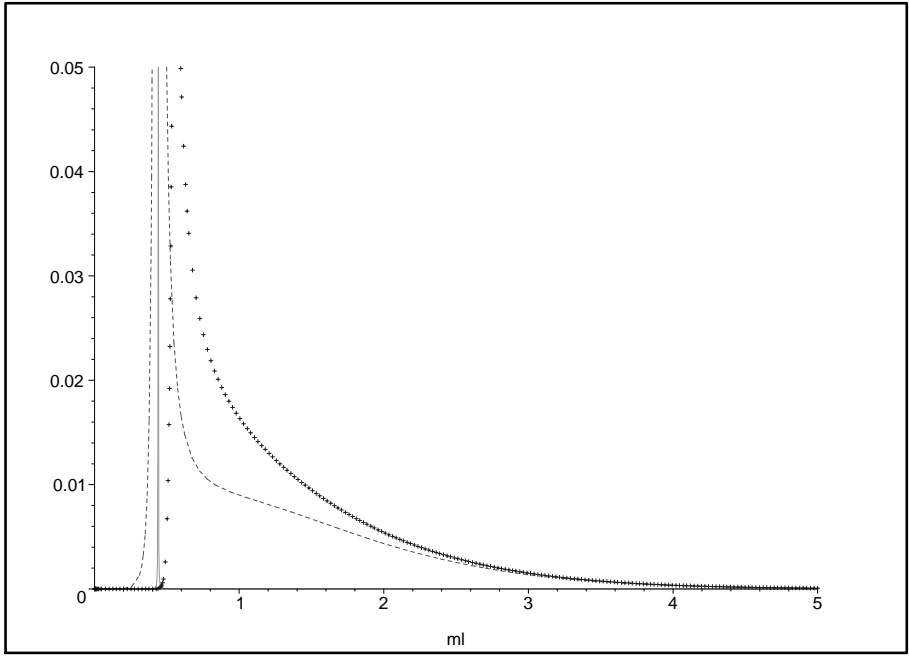}
\caption{\label{fig:IntFuerza} The figure shows the contribution
 to the Casimir force of the continuous modes
($n=0$) without approximations, namely the integrand of eq.
(\ref{fT}), as a function of mass in units of the separation between
plates $ml$. The value $m_6 l=\sqrt{5/8}$ is used for the scalar
field mass. The curve in crosses represents the case $\kappa
l=1/50$, the dashed one is associated with $\kappa l=1$ whereas the
continuous one corresponds to $\kappa l=50$.}
\end{figure}

At this point eq. (\ref{fT}) for $f_{T}$ does not seem to reproduce
the standard 4D Casimir force in particular due to the fact that
$\psi_m$ are not true localized states. Nevertheless by alluding to
the existence of a quasilocalized resonant mode and a light mode
sector given by (\ref{phiappox}) there are two contributions giving
the following effective Casimir force
\begin{eqnarray}
f_{\mathrm{eff}}&=&\frac{3}{2}f_{4D}(m_q)+f_{\mathrm{light}},\label{efectiva} \\
f_{\mathrm{light}}&: =& \frac{1}{\pi \kappa^3 m_q^4}\int_{m\, \ll \,
m_q} m^6\, f_{4D}(m)\, dm. \nonumber
\end{eqnarray}
This idea is also clear from Fig. \ref{fig:IntFuerza} showing
conditions under which the quasilocalized or resonant mode
dominates. The physical interpretation now is obvious. The first
term corresponds to 3/2 of the standard 4D Casimir force of a
massive scalar field whose mass is $m_q$, i.e., we regain the
functional form of the standard Casimir effect up to a numerical
factor. The second term accounts for the effect of the light modes
in the low energy regime: $m\ll R^{-1}, \kappa$, which imply $n=0$.
Clearly this result is connected to the form the Green's function
gets in the light mode approximation as presented in eq.
(\ref{Green'sFContr6DMassive}). Moreover we can evaluate numerically
an approximation of this integral noticing that for large $m$: $m\gg
l^{-1}$, the standard Casimir force $f_{4D}(m)$ vanishes
exponentially \cite{Milton:2001yy}. This fact allows the extension
of the integration interval in eq. (\ref{efectiva}) from $\int_{m\,
\ll \, m_q} \rightarrow \int_0^\infty$, obtaining
\begin{equation}
f_{\mathrm{light}} \approx \frac{259.5}{\pi^5 (\kappa \, l)^3 (m_q
\,l)^4}f_{4D}(0)
\end{equation}
where we have done the change $m = \alpha /l$ and evaluated
numerically the integral $ \int_0^\infty \alpha^6 f_{4D}(\alpha)
d\alpha = 17.3$. $f_{4D}(0)$ represents the standard Casimir force
of a massless scalar field. This approximation is the same one that
is realized for the scalar potential in \cite{Dubovsky:2000am}.
Recalling the light scalar field approximation $m_q l\ll 1
\Rightarrow f_{4D}(m_q) \approx f_{4D}(0)$ and considering
$f_{\mathrm{light}}\ll \frac{3}{2} f_{4D}$ in eq. (\ref{efectiva})
one gets the lower bound $(\kappa \, l)^3 (m_q \,l)^4\gg 1$.

\section{Massless scalar field}\label{masslessstringlike}

In this section we describe how the mass spectrum for the scalar
field changes when its bulk mass $m_4$ is zero. The basic difference
is that whereas for the massive case there does not exist a zero
mode but a quuasilocalized or resonant one, for the massless case
there does exist a zero mode state. We shall also determine the
Casimir force for the standard setting of two parallel plates.

\subsection{Green's function}

Let us start by considering the field equation (\ref{ecu:field}) by
setting $m_6=0$. One can keep track of this condition resulting
again in (\ref{Thetaeq}) and (\ref{varphieq}) but changing eq.
(\ref{Yeq}) for $\psi(y)$ which naturally leads to
\begin{equation}\label{ecforPsiMassless6D}
\left(\partial_{y}^{2}-5\kappa\,sgn(y)\partial_{y}+m^2\,e^{2\kappa|y|}\right)\psi
=  0,
\end{equation}
and is equivalent, after the change of variable $\tilde y=e^{\kappa
y}/\kappa$ and $ \tilde \psi(\tilde y)=\tilde y^{5/2}\psi(y)$ to
\begin{equation}\label{ecumodos1}
\partial_{\tilde y}^2\tilde \psi+\frac{1}{\tilde y}\partial_{\tilde y}\tilde \psi+
\left(m^2-\frac{\gamma_0^2}{\tilde y^2}\right)\tilde \psi=0,
\end{equation}
where $\gamma_0=\frac{5}{2}$. Mathematically the fact that $\gamma$
be a rational number is the feature that allows to have a zero mode
solution. Notice that this equation can be obtained from
(\ref{ecforPsi}) by setting $m_6=0$ so that
$\gamma|_{m_6=0}=\gamma_0$. When $m=0$ eq. (\ref{ecumodos1})
includes as a solution
\begin{equation}
\psi_0(y)=\sqrt{\frac{3}{2}\kappa}\, , \quad \mbox{which
satisfies}\quad \int_{-\infty}^{\infty}dye^{-3\kappa |y|}\psi_0^2=1.
\end{equation}
When $m>0$, the normalized eigenstates are
\begin{equation}\label{psiy1}
\psi_m(y)=e^{\frac{5}{2}\kappa y}\,\sqrt{\frac{m}{2\kappa}}\left[
a_mJ_{\gamma_0}\left(\frac{me^{\kappa y}}{\kappa}\right)+ b_m
N_{\gamma_0}\left(\frac{me^{\kappa y}}{\kappa}\right) \right]\, ,
\end{equation}
where the constants are given by
\begin{equation}
a_m=-\frac{A_m}{\sqrt{1+A_m^2}},\quad b_m=\frac{1}{\sqrt{1+A_m^2}},
\hspace{0.5cm} \mbox{and} \hspace{0.5cm} A_m=\frac{
N_{\gamma_0-1}\left(\frac{m}{\kappa}\right)}{J_{\gamma_0-1}\left(\frac{m}{\kappa}\right)}
\,,
\end{equation}
as can be shown by recalling that all these solutions satisfy the
same normalization and boundary conditions as in the massive case
$m_6\neq 0$, i.e. (\ref{normalizationmy})-(\ref{Neumann}). The
existence of the zero mode $\psi_0$ associated to the noncompact
coordinate $y$ is the main difference between the massless and the
massive scalar field cases. This result implies that, in the
massless $\Phi$ case, the contribution of the noncompact extra
dimension to the 4D mass includes as a possible value
$m_4^2=(n/R)^2$ in contrast with the massive case. In particular we
have a state corresponding to $m_4=m=n=0$, which is a true localized
massless state. See Fig. \ref{fig:modos} for an illustration of this
fact.

Taking into account all the ingredients together we are now in
position to write down the 6D Green's function, which in this case
satisfies the equation (\ref{green1}) with $m_6=0$ and its
expression in eigenfunctions, analogue of (\ref{6DComGreen}),
becomes
\begin{equation}\label{GreenMassles6D}
G_{6D}^{\, 0}=\sum_{n} \Theta_n(\theta) \Theta_n (\theta') \left[
\frac{1}{\kappa}\psi_0(y)\psi_0(y')+ \int \frac{dm}{\kappa} \psi_{
m}(y)\psi_{m}(y')\right] G_{4D}\left(x,x';m_4\right).
\end{equation}
Clearly one can read this full Green's function as made up of
individual massive modes' 4D Green's functions weighted by the
modes' wave functions. In particular in this massless $\Phi$ case
there is a localized zero mode contribution as opposed to the
massive case.

In the light modes approximation $m\ll R^{-1}, \kappa$, hence $n=0$,
the effective $4D$ Green's function is
\begin{equation}\label{Green'sModesMassles6D}
G_{\mathrm{eff}}^{\, 0} (x,x')\approx \frac{1}{2\pi
R}\frac{1}{\kappa}\, \psi_{0}^2(0)\, G_{4D}(x,x';m_4=0) +
\frac{1}{2\pi R} \int_{m \, \ll \kappa} \frac{dm}{\kappa} \,
\psi_m^2(0) \, G_{4D}(x,x';m_4),
\end{equation}
where we have taken the limit $y,y'\rightarrow 0$, which gives the
physics on the brane.

The first term corresponds to the zero mode of the masless scalar
field $\Phi$ and is proportional to the 4D Green's function of a
massless scalar field. Consequently this term is associated with
standard 4D physics. The second term is the contribution of the
massive modes of the continuous tower of states.

\subsection{Casimir force}

The total Casimir force between plates in the string like defect can
be computed from (\ref{GreenMassles6D}) following the same procedure
as in the previous section only extended by the presence of the zero
mode. In this manner one obtains
\begin{equation} \label{fTmassless}
f_T^{\, 0}= \sum_{n}\left( \frac{\psi_0^2(0)}{\kappa}+ \int_{m \,
\ll \, \kappa}\frac{dm}{\kappa}\,\psi_m^2(0)\right)
f_{4D}\left(m_4\right) .
\end{equation}
Notice that although the second term for the light modes here is
formally the same as the one in the massive scalar field case
discussed in the previous section, they truly differ due to the fact
that in that case $\psi_m^2 \stackrel{m_6\neq 0}{\sim} m^6$, whereas
here $\psi_m^2 \stackrel{m_6=0}{\sim} m^2$. Explicitly,  in the
light modes approximation $m \ll R^{-1},\kappa$
\begin{equation}
A_m\approx -3\left(\frac{m}{\kappa}\right)^{-3} \hspace{0.5cm}
\mbox{and} \hspace{0.5cm} \psi_m^2(0)\approx \frac{m^2}{\pi
\kappa^2},
\end{equation}
resulting in an effective Casimir force
\begin{eqnarray}
f_{\mathrm{eff}}^{\, 0}&\approx&\frac{3}{2}\,
f_{4D}(0)+f_{\mathrm{light}}^{\, 0}. \label{efectivacero} \\
f_{\mathrm{light}}^{\, 0}&:=&\frac{1}{\pi \kappa^3}\int_{m \, \ll \,
\kappa} m^2 f_{4D}(m)\ dm. \nonumber
\end{eqnarray}
The light modes contribution can be evaluated further as in the
massive case obtaining finally
\begin{equation}
f_{\mathrm{light}}^{\, 0}\approx \frac{2.55}{\pi^5 (\kappa \,
l)^{3}}\, f_{4D}(0).
\end{equation}
Considering again $f_{\mathrm{light}}^{\, 0}\ll
\frac{3}{2}f_{4D}(0)$ in (\ref{efectivacero}) one gets the lower
bound $\kappa l \gg 10^{-1}$. By taking $l\sim 10^{-6}$m of typical
Casimir experiments one gets an upper bound for the anti de Sitter
radius of $\kappa^{-1}\ll 10^{-5}$m.

\section{Discussion}\label{secdiscussion}

To avoid conflict with well tested low energy physics brane world
models are usually built up accordingly, namely, they include a big
enough mass gap of a given field separating the zero mode, which
yields standard 4D physics, from a continuum sector of massive modes
in the case of non compact extra dimensions, which produce
corrections to 4D physics. In this way physical effects are
investigated mostly in the high energy regime including also
astrophysics or cosmology. Interestingly, just for the same reason
submillimeter experiments of Newton's gravity law
\cite{Hoyle:2000cv,Kapner:2006si} probe the mode structure of
gravity in the low energy regime for brane worlds the case for
matter fields can be raised in relation with precision tests like
the Casimir force or other atomic experiments
\cite{Lamoreaux:2005gf,Bluhm:2000tv,Luo:2006ck,MoralesTecotl:2006eh}.

As a specific model proposal of low energy test in this paper we
consider the Casimir force produced by a scalar field with and
without mass in a Randall-Sundrum type of brane world consisting of
a single brane extended by one compact extra dimension or RSII-1.
Such a higher dimensional scenario is interesting because it allows
localization of gauge fields \cite{Dubovsky:2000am}. We adopt the
standard setting of the Casimir effect which involves two parallel
flat plates on which the scalar field is subject to Dirichlet
boundary conditions. The plates lie along the single brane of the
RSII-1 model.

To calculate the Casimir force we use the Green's function method.
The first result that makes a crucial difference between the massive
and massless scalar field is related to the mass spectrum they
present. Whilst both include a continuous sector the former lacks a
zero (massless) mode possessing instead a quasilocalized or resonant
mode, see eq. (\ref{phiappox}). The latter case contains a zero
mode. The Green's function and the Casimir force it produces depend
in detail on this fact. Nonetheless, and this is our second result,
they both turn out to be expressed, in the low energy regime, as the
combination of the individual modes acting as in 4D but weighted by
the values of the modes' wave functions on the brane as shown in
eqs. (\ref{Green'sFContr6DMassive}), (\ref{fT}),
(\ref{Green'sModesMassles6D}) and (\ref{fTmassless}). Investigating
the dependence of the correction terms to the standard Casimir force
led us to  eqs. (\ref{efectiva}) and (\ref{efectivacero}), for the
massive and massless cases, respectively. By looking at the
conditions under which such corrections do not dominate standard
Casimir force produced to kinds of bounds. In the massive scalar
field case we got the lower limit of the product  $(\kappa l)^3 (m_q
l)^4 \gg 1$ whereas in the massless case the upper bound for the
anti de Sitter radius $\kappa^{-1}\ll 10^{-1}$m. Incidentally this
limit turns out to be weaker than other obtained based upon the Lamb
shift for Hydrogen in a similar brane world
\cite{MoralesTecotl:2006eh}.

We should keep in mind our model proposal is too simple at present
to be compared with actual experimental data. Indeed we are giving
only first steps in this direction to fill in the gap for low energy
tests of brane world models in current literature.

As a further development it would be interesting to compare the
analysis following dimensional regularization like in refs.
\cite{Poppenhaeger:2003es,Linares:2005cj,Frank:2007jb} with those
adopting Green's function techniques like in the present paper.
Moreover it seems possible to generalize our results to an arbitrary
number of extra compact dimensions or RSII-p and it should be
possible to study the mass spectrum as well as Greens' function and
the corresponding Casimir force \cite{Linares:2008}.

\begin{acknowledgments}
We are grateful to Kim Milton for enlightening correspondence. This
work was partially supported by Mexico's National Council of Science
and Technology (CONACyT), under grants (SEP-CONACyT)-2004-C01-47597
and (SEP-CONACyT) CB-2005-C01-51132-F. The work of O.P. was
supported by CONACyT Scholarship number 162767 and by a scholarship
of the grant (SEP-CONACyT)-2004-C01-47597.
\end{acknowledgments}

\bibliography{casimir-scalar}

\end{document}